\documentclass[apjl]{emulateapj}
\usepackage{amstext}
\usepackage{apjfonts}
\usepackage{amsmath}
\usepackage{amssymb}
\usepackage[colorlinks=true,linkcolor=blue,citecolor=blue]{hyperref}
\usepackage{comment}
\usepackage{mathrsfs}

\newcommand{\beq}{\begin{equation}}
\newcommand{\eeq}{\end{equation}}

\begin{document}

\title{On the Assembly Rate of Highly Eccentric Binary Black Hole Mergers} 
\author{Johan Samsing$^{1,2}$ and Enrico Ramirez-Ruiz$^{3}$} 
\altaffiltext{1}{Department of Astrophysical Sciences, Princeton University, Peyton Hall, 4 Ivy Lane, Princeton, NJ 08544, USA}
\altaffiltext{2}{Einstein Fellow}
\altaffiltext{3}{Department of Astronomy and Astrophysics, University of California, Santa Cruz, CA 95064, USA}

%%%%%%%%%%%%%%%%%%%%%%%%%%%%%%%%%%%%%%%%%%%%%%%%%%%%%%%%%%%%%%%%%%%%%%%%   
\begin{abstract} 
%%%%%%%%%%%%%%%%%%%%%%%%%%%%%%%%%%%%%%%%%%%%%%%%%%%%%%%%%%%%%%%%%%%%%%%%

In this {\it Letter} we calculate the fraction of highly eccentric  binary black hole (BBH) mergers resulting from  binary-single interactions.
Using an $N$-body code that includes post-Newtonian correction terms, we show that $\gtrsim 1\%$ of all BBH mergers resulting from this channel
will have an eccentricity $e>0.1$ when coming into the LIGO frequency band. As the majority of BBH mergers forming in globular clusters
are assembled through three-body encounters, we suggest that such interactions are likely to dominate
the population of high eccentricity BBH mergers detectable by LIGO. The relative  frequency of highly eccentric events could eventually help to constrain the astrophysical origin of BBH mergers. 
\end{abstract}
%%%%%%%%%%%%%%%%%%%%%%%%%%%%%%%%%%%%%%%%%%%%%%%%%%%%%%%%%%%%%%%%%%%%%%%%

\section{Introduction}

Recent studies indicate that there are many mechanisms  to assemble  binary black hole
(BBH) binaries,  with merger  rates that  are broadly consistent with those  currently inferred from LIGO observations \citep{2016PhRvL.116f1102A}.
Such models include BBH mergers forming as a result of isolated binary evolution \citep[e.g.][]{2016Natur.534..512B},
dynamical interactions in globular
clusters (GCs) \citep[e.g.][]{2015PhRvL.115e1101R, 2016PhRvD..93h4029R, 2016ApJ...824L...8R}, single-single captures
in galactic nuclei \citep[e.g.][]{2009MNRAS.395.2127O}, gas hardening of BBH binaries in active galactic nuclei disks
\citep[e.g.][]{2017ApJ...835..165B,  2017MNRAS.464..946S, 2017arXiv170207818M}, as well as
primordial BBH mergers \citep[e.g.][]{2016PhRvL.116t1301B, 2016PhRvD..94h3504C}.
In addition, not only are the rates estimated to be of similar magnitude, but their associated BBH mass distributions can be similarly constructed  as well. As a result, it might be difficult  to discern  useful astrophysical information from future gravitational wave (GW) data if only merger rates and BH masses are
measured and recorded \citep[e.g.][]{2017arXiv170208479C}.

Despite some similarities, the distribution
of BBH eccentricity and BH spin orientations at the time of merger, are likely to differ between models.
For example, one expects spins to be mostly aligned and eccentricity to be near zero for BBH mergers
forming from isolated binary channels \citep[e.g.][]{2000ApJ...541..319K}, where dynamical channels
are more likely to lead to randomized spin orientations and non-zero
eccentricities \citep[e.g.][]{2006ApJ...640..156G, 2009MNRAS.395.2127O, 2014ApJ...784...71S, 2016ApJ...832L...2R}.
Estimating the BBH eccentricities and spins expected  from different assembly models could thus provide us with key discerning information.
Hence it is not surprising that major effort is currently being devoted to develop templates and
fast techniques for measuring GW waveforms for varying spin and eccentricity \citep[e.g.][]{2016PhRvD..94b4012H, 2016arXiv160905933H}.

In this {\it Letter} we study the eccentricity distribution of BBH mergers resulting from  dynamical
binary-single interactions \citep{Hut:1983js, 2014ApJ...784...71S, 2016arXiv160909114S} in GCs \citep{Heggie:1975uy}.
As  shown in \citet{2014ApJ...784...71S}, when
post-Newtonian (PN) corrections \citep[e.g.][]{Blanchet:2006kp} are implemented in the few-body equation-of-motion (EOM), 
BBH mergers can occur during the chaotic three-body interaction \citep[e.g.][]{2006ApJ...640..156G}. We here refer to such
BBH mergers as {\it GW inspirals}.
Interestingly, GW inspirals often come into the LIGO frequency band with a very high eccentricity.  Here we perform binary-single experiments with PN terms implemented in the EOM, from which we reconstruct  the
eccentricity distribution of BBH mergers found by \cite{2016PhRvD..93h4029R}, and, for the first time,  provide an estimate for the number of mergers taking place at  high eccentricities. The above exercise leads us to conclude  that $\gtrsim 1\%$ of all BBH mergers
forming in GCs via the binary-single channel  have eccentricities $e>0.1$ when entering the LIGO band.

\section{Black Hole Binary-Single Interactions}
Interactions between a binary and a single interloper are among the most common few-body interactions taking place in  a typical GC.
Such encounters  not only play a key dynamical role \citep{Heggie:1975uy},
but might be essential  for assembling  BBHs as well. As an example,
recent work by \citet{2016PhRvD..93h4029R} estimates that the vast majority of BBHs that are ejected from GCs
and subsequently merge through GW emission, originate from binary-single
interactions. 

Motivated by this, we derive here the eccentricity distribution of BBH mergers
resulting from binary-single interactions when PN terms are included in
the three-body EOM. We only consider binary-single interactions for which the total initial
energy is negative, a limit that is usually referred to as the hard binary (HB) limit \citep[e.g.][]{Hut:1983js}.

\begin{figure}
\centering
\includegraphics[width=\columnwidth]{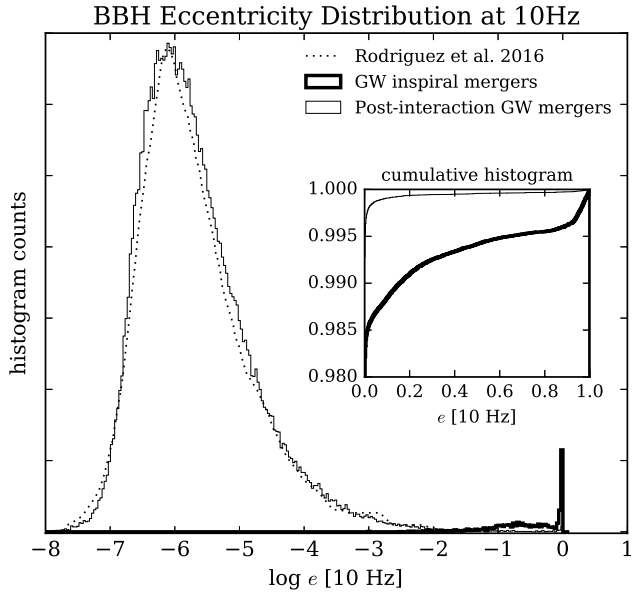}
\caption{Results from $2.5 \times 10^{5}$ binary-single interactions between [BH($20M_{\odot}$), BH($20M_{\odot}$)] binaries
with an initial SMA distribution that is uniform in $\log(a_{0})$ between $0.1-0.2$ AU
and an incoming single BH($20M_{\odot}$) with a velocity at infinity $v_{\infty} = 10$ kms$^{-1}$.
The $N$-body code used for these scatterings includes GW emission at the $2.5$PN level.
Shown are the eccentricity distributions of all assembled BBH mergers at the time their GW frequency is $10$Hz. The {\it thin solid} line (peaking at around $e \approx 10^{-6}$) shows the eccentricity distribution for all
post-interaction binary GW mergers, while the {\it thick solid} line (peaking at around $e \approx 1$) shows the eccentricity distribution of all GW inspiral
mergers that form during the chaotic three-body interaction. As seen, the BBHs that merge during the chaotic three-body interaction
have a very high eccentricity when they come into the LIGO band. This population only shows
up when PN corrections are consistently included in the EOM \citep{2006ApJ...640..156G}.
The {\it insert plot} shows the corresponding cumulative distributions, where the {\it thick solid} line now shows results for
the joined population of post-interaction binary GW mergers and GW inspiral mergers. From this line we see
that about $1\%$ of all BBH mergers have a high eccentricity at merger. We suggest that such interactions
might dominate  the production rate of  high eccentricity BBH mergers. 
The {\it dotted line} shows the results from \citet{2016PhRvD..93h4029R}, which are based on
detailed GC models. The high eccentricity tail is  not seen in their study, as  PN effects are not included.}
\label{fig:BBH_ecc_with25PN}
\end{figure}

\subsection{Post-Interaction BBH Mergers}
The most likely outcome of a binary-single interaction
is a binary with an unbound single \citep{Heggie:1975uy}. In this study
we generally refer to the final binary from that outcome as the {\it post-interaction binary}.
Such post-interaction binaries are generally formed with an eccentricity distribution that to leading order
follows a thermal distribution, and binding energies that are slightly higher than that of the initial
target binary \citep{Heggie:1975uy}.
As a result, the GW merger time of post-interaction binaries
can be significantly shorter than the GW merger time of the initial target binary \citep[][]{Peters:1964bc}.
For this reason, the binary-single channel is considered to be an effective pathway for transforming binaries with long GW inspiral times,
into binaries that could merge within a Hubble time.

The rate of BBH mergers forming in this way has recently
been studied by \citet{2015PhRvL.115e1101R, 2016PhRvD..93h4029R} using detailed dynamical simulations of GCs. The rate of BBH mergers forming in GCs  is estimated by these authors  to be somewhere between $1-100$ per Gpc$^{-3}$ yr$^{-1}$, with all  their resulting mergers having circularized long before
entering the LIGO band. The population of post-interaction BBHs
is thus not expected to contribute to the  rate of highly eccentric BBH mergers.

\subsection{Formation of Highly Eccentric BBH Mergers}
High eccentricity GW inspirals have recently been shown to be produced  during resonant binary-single
interactions when PN terms are included in the three-body EOM \citep{2006ApJ...640..156G, 2014ApJ...784...71S, 2016arXiv160909114S}.
As described  in \citet{2014ApJ...784...71S, 2016arXiv160909114S}, a resonant interaction can be envisioned as
a series of intermediate states that are each composed of a binary
with a bound single. Between each of these states, the three objects undergo a strong interaction that results in
a semi-random redistribution of orbital energy and angular momentum.
Each intermediate state binary therefore has a finite probability for being formed with a high eccentricity, even
if the initial target binary is circular. If the eccentricity is high enough,
the intermediate state binary will undergo a GW inspiral and merge while still being bound to the single object.
On the other hand, if the eccentricity is too low, the corresponding GW inspiral
time is too long, and the single will simply return to the intermediate state binary after which a new binary-single state will form.
As a result, GW inspirals will generally form at high eccentricity.

Although such GW inspirals are generally assembled with an initial high eccentricity, they do not necessarily come into the LIGO frequency
band with the same  high eccentricity. In fact, as
described in \citet{2014ApJ...784...71S}, a finite fraction of the GW inspirals will circularize before entering the LIGO band.
This population originates from the
fraction of interactions where the bound single in the intermediate state is send out on an orbit with close to zero
orbital energy. For this reason, it is important to not only identify and count the number of GW inspirals, but also to evolve them
 until they reach the frequency band observable by LIGO. We thus now turn our attention to study the eccentricity distribution of BBH mergers forming through the binary-single channel
at   the time their GW  frequency is $10$Hz. We refer to this frequency threshold as {\it coming into the LIGO band}. 
As inferred from our previous arguments, the eccentricity distribution is expected to have two peaks.
One arising  from the post-interaction binary GW mergers and another one  from the GW inspirals forming during the three-body interaction.

For this study, we perform $2.5\times 10^{5}$ numerical binary-single scatterings with $2.5$PN corrections
included in the EOM \citep[for details on our $N$-body code, the reader is referred to][]{2016arXiv160909114S}.
The scatterings are between [BH($20M_{\odot}$), BH($20M_{\odot}$)] binaries with an initial
semi-major axis (SMA) distribution that is uniform in $\log(a_{0})$, where $a_{0}$ denotes the initial SMA, between $0.1-0.2$ AU,
and an incoming single BH($20M_{\odot}$). The relative velocity at infinity is set to $v_{\infty} =10$ kms$^{-1}$,
however, our presented relative rate estimates do not depend on this value as long as we are in the HB limit \citep{2014ApJ...784...71S}.
These initial conditions (ICs) are inspired by the derived distributions of BBH mergers in our local universe from \cite{2016PhRvD..93h4029R}.
For each BBH that forms and eventually merges we follow its SMA and eccentricity using the GW quadrupole
equations given in \citet{Peters:1964bc}, together with its GW peak frequency derived using the fitting formulae by \citet{Wen:2003bu}.
We only include BBHs that merge within a Hubble time, and for the post-interaction binaries we further require that the
kick velocity $> 50$ kms$^{-1}$. These two requirements are included for consistency, but do not significantly affect
our results.

Our scattering results are shown in Figure \ref{fig:BBH_ecc_with25PN}. We first note that our derived post-interaction binary eccentricity distribution
at $10$Hz is very similar to that obtained  by \cite{2016PhRvD..93h4029R} using a more sophisticated
GC modeling approach. While this  is expected due to our choice of ICs, it gives credence to the validity of the formalism used here.
We note that the main shape of the distribution depicted  in Figure \ref{fig:BBH_ecc_with25PN} originates from the thermal distribution of eccentricities \citep{Heggie:1975uy}, where
both  the BBH mass and initial SMA determine the location of the peak.
The high eccentricity component  of the  distribution, originating from our inclusion of GW emission in the EOM through the $2.5$PN term, is clearly evident  in Figure \ref{fig:BBH_ecc_with25PN} and contains  about $\gtrsim 1\%$ of all BBH mergers. If $2.5$PN corrections are not included, this high eccentricity component  within the LIGO
frequency band is absent from the distribution.

Having numerically derived the relative fraction of highly eccentric mergers, it is important to explore how this fraction changes when the SMA $a_{0}$ and  BH mass
$m_{\rm BH}$ are altered. This  can be estimated analytically for when the merger time of the initial target binary, $\tau_{0}$, is either
shorter or longer than the Hubble time, $t_{\rm H}$. Both limits are important as $\tau_{0} \approx t_{\rm H}$ for the selected  ICs.
We start by noting that low eccentricity mergers are dominated by post-interaction binary mergers,
while high eccentricity ones are dominated by GW inspirals. The relative rate of highly eccentric BBH mergers
can thus be approximated by the ratio between the GW inspiral cross section and the cross section for post-interaction binaries
that merge within a Hubble time. In the limit where $\tau_{0} < t_{\rm H}$, the
cross section for post-interaction binaries that merge within a Hubble time, $\sigma_{\rm M}$, scales simply as the gravitational
focusing term $\sigma_{\rm M} \propto a_{0}m_{\rm BH}$. In the limit where $\tau_{0} > t_{\rm H}$, one finds that the cross section scales
instead as $\sigma_{\rm M} \propto a_{0}^{-1/7}m_{\rm BH}^{13/7}$, where we have assumed that
the post-interaction binary eccentricity distribution follows a thermal distribution $P(e) = 2e$ \citep{Heggie:1975uy}.
The GW inspiral cross section, $\sigma_{\rm I}$, was shown in \citet{2014ApJ...784...71S} to scale as
$\sigma_{\rm I} \propto a_{0}^{2/7}m_{\rm BH}^{12/7}$.
The fraction of BBHs that merge
with relative high eccentricity, $f_{\rm e\approx 1}$, can  then be written as 
\begin{equation}
f_{\rm e\approx 1} \propto {a_{0}^{-5/7}}{m_{\rm BH}^{5/7}},\  \tau_{0} < t_{\rm H},
\end{equation}
\begin{equation}
f_{\rm e\approx 1} \propto {a_{0}^{3/7}}{m_{\rm BH}^{-1/7}},\  \tau_{0} > t_{\rm H}.
\end{equation}
Say we keep $m_{\rm BH}$ fixed, we then see that $f_{\rm e\approx 1}$ will always increase when $a_{0}$ is varied, provided that $\tau_{0} \approx t_{\rm H}$. As $\tau_{0} \approx t_{\rm H}$ for our ICs, we can therefore conclude that any large changes in $a_{0}$ will  lead to a larger $f_{\rm e\approx 1}$ than the one reported in this {\it Letter}.
Our scalings essentially show that $f_{\rm e\approx 1}$ is not expected
to change much for variations similar to those  found by \citet{2016PhRvD..93h4029R}. The fraction of high eccentricity BBH mergers derived here, is
likely to be an accurate estimate of the fraction  for a broad  range of GC properties.

\section{Discussion}

Recent studies show that the BBH spin vectors are likely to differ between different merger channels
\citep[e.g.][]{2016ApJ...832L...2R, 2016MNRAS.462..844K, 2017arXiv170200885Z},
however, only little work has been done on deriving corresponding differences in BBH eccentricity
despite its importance \citep[e.g.][]{2016MNRAS.458.3075A, 2017arXiv170208479C}. 
In this {\it Letter} we have shown that $\gtrsim 1\%$ of all BBH mergers forming in GCs via binary-single
interactions will have eccentricities $e>0.1$ when coming into the LIGO band.
Together with spin, eccentricity is probably the other most promising parameter
to help observationally distinguish between different astrophysical  BBH merger channels.

In \cite{2016ApJ...816...65A} it was argued that the rate of high eccentricity ($e > 0.1$) BBH mergers observable with Advanced LIGO,
is likely to be dominated by BBHs that are driven to merger  via Lidov-Kozai (LK)
oscillations \citep{1962P&SS....9..719L, 1962AJ.....67..591K} in hierarchical triples formed in GCs.
By  using a MC method  for simulating the evolution of GCs \citep{2015ApJ...800....9M} together  with
the ARCHAIN code \citep{2008AJ....135.2398M}, \citet{2016ApJ...816...65A}
concluded that $\approx 1\%$ of all BBH mergers formed in GCs will merge with high eccentricity via the LK mechanism.

\citet{2016ApJ...816...65A} compared their derived LK merger rates to a broad range of
other dynamical channels, from which they concluded that the LK channel is likely to dominate the population of
eccentric BBH mergers. However, they overlooked the possibility of forming a sizable number of high eccentricity BBH GW
mergers through the binary-single channel.
\citet{2016ApJ...816...65A} did note that about $1\%$ of all BBH mergers assembled through the binary-single channel
are likely to merge at high eccentricity \citep[as argued by][]{2014ApJ...784...71S}, but they mistakingly estimated the corresponding   rate by using their {\it In-cluster mergers} population \citep[see Table 1 in][]{2016ApJ...816...65A}. A more accurate estimate can be made  when considering  their {\it Ejected binaries} population, as  this group primarily  represents  the hard (unbound)  binaries formed via binary-single interactions \citep{2016PhRvD..93h4029R}.
Using the population  of {\it Ejected binaries}  in \citet{2016ApJ...816...65A} one finds that similar rates of highly eccentric GW mergers can be  derived from binary-single interactions than  from their proposed KL channel.

In this study we have not only demonstrated that a significant number of GW inspiral mergers do form in three-body interactions,
but also that they keep their high eccentricity until they come into the LIGO frequency band. While deriving an absolute merger rate is  difficult
our results robustly  indicate that the binary-single channel is a competitive alternative  for producing high
eccentricity GW mergers. In fact, the rate of LK mergers estimated by \citet{2016ApJ...816...65A} is  highly uncertain,
as their sample of hierarchical triples was not self-consistently evolved in their  simulations. The rate derived here for GW inspirals  is also not exact 
as we have only considered equal mass isolated binary-single interactions. Having said this, we do find that GW inspiral mergers
happen relative promptly, are easy to identify and their outcomes are only mildly  dependent on black hole mass and initial SMA distribution. This indicates
that a  relative rate of highly eccentric GW mergers can be robustly estimated from this dynamical channel.

\acknowledgments{
We thank the Aspen Center for Physics for its hospitality while part of this work was completed and acknowledge C. Rodriguez and M. Macleod for helpful discussions.
Support for this work was provided by  the David and Lucile Packard Foundation, UCMEXUS (CN-12-578) and  NASA through an Einstein
Postdoctoral Fellowship grant number PF4-150127, awarded
by the Chandra X-ray Center, which is operated by the
Smithsonian Astrophysical Observatory for NASA under
contract NAS8-03060.
}

%%%%%%%%%%%%%%%%%%%%%%%%%%%%%%%%%%%%%%%%%%%%%%%%%%%%%%%%%%%%%%%%%%%%%%%%   
%\bibliography{NbodyTides_papers}
\bibliographystyle{apj}

%%%%%%%%%%%%%%%%%%%%%%%%%%%%%%%%%%%%%%%%%%%%%%%%%%%%%%%%%%%%%%%%%%%%%%%%   

\end{document}